\pdfoutput=1

\documentclass[11pt]{article}

\usepackage[final]{acl}

\usepackage{times}
\usepackage{latexsym}

\usepackage{amsmath}
\usepackage{multirow}
\usepackage{booktabs}
\usepackage{makecell}
\usepackage{color,soul}
\usepackage{xcolor}
\usepackage{tcolorbox}

\usepackage[T1]{fontenc}

\usepackage[utf8]{inputenc}

\usepackage{microtype}

\usepackage{inconsolata}

\usepackage{graphicx}

\title{Evaluating Creativity and Deception in Large Language Models: A Simulation Framework for Multi-Agent Balderdash}

\author{Parsa Hejabi\thanks{Equal contribution.}\textsuperscript{\textdagger} \And Elnaz Rahmati\footnotemark[1]\textsuperscript{\textdagger} \And Alireza S. Ziabari\textsuperscript{\textdagger} \And Preni Golazizian\textsuperscript{\textdagger} \AND Jesse Thomason\textsuperscript{\textdagger} \And Morteza Dehghani\textsuperscript{\textdagger} \AND
        \normalfont \textsuperscript{\textdagger} \small University of Southern California \\
        \texttt{\small\{hejabi, erahmati, salkhord, golazizi, jessetho, mdehghan\}@usc.edu}}

\begin{document}
\maketitle
\begin{abstract}
Large Language Models (LLMs) have shown impressive capabilities in complex tasks and interactive environments, yet their creativity remains underexplored. This paper introduces a simulation framework utilizing the game Balderdash to evaluate both the creativity and logical reasoning of LLMs. In Balderdash, players generate fictitious definitions for obscure terms to deceive others while identifying correct definitions. Our framework enables multiple LLM agents to participate in this game, assessing their ability to produce plausible definitions and strategize based on game rules and history. We implemented a centralized game engine featuring various LLMs as participants and a judge LLM to evaluate semantic equivalence. Through a series of experiments, we analyzed the performance of different LLMs, examining metrics such as True Definition Ratio, Deception Ratio, and Correct Guess Ratio. The results provide insights into the creative and deceptive capabilities of LLMs, highlighting their strengths and areas for improvement. Specifically, the study reveals that infrequent vocabulary in LLMs' input leads to poor reasoning on game rules and historical context.\footnote{\url{https://github.com/ParsaHejabi/Simulation-Framework-for-Multi-Agent-Balderdash}}
\end{abstract}

\section{Introduction}
Large Language Models (LLMs) have recently been employed as agents in various complex tasks, showcasing their potential in dynamic, interactive environments \citep{10373065,singh2024twostep}. This has led to a growing interest in LLM-based multi-agent systems (LLM-MA), particularly within the realm of gaming \citep{mukobi2024welfare,xu2023exploring}. Games offer a structured yet flexible platform to analyze and understand LLM behavior under diverse scenarios \citep{light2023avalonbench}.

Currently, LLMs are typically evaluated through static tasks \citep{lee2023natural,zhao2024assessing,gomez-rodriguez-williams-2023-confederacy}. Traditional games like Avalon \citep{wang2023avalons} and Werewolf \citep{xu2024language} have also been used to benchmark LLMs, focusing on logical reasoning and strategic interaction. These games require players to engage in deception, deduction, and negotiation, providing valuable insights into LLMs' decision-making processes. However, these studies often overlook the assessment of creativity.

To address this gap, we introduce a simulation framework for the game \textit{Balderdash}. In this game, players generate plausible yet fictitious definitions for obscure terms, aiming to deceive other players while identifying the correct definitions. 
We argue that Balderdash can be used to evaluate both the creativity and logical reasoning of LLMs, challenging the models to balance these two crucial aspects and providing a comprehensive assessment of their capabilities.


In this paper, we aim to assess the creativity of LLMs by evaluating their ability to generate plausible definitions for obscure words in Balderdash. We will further examine their logical reasoning skills by observing how effectively they deceive opponents and identify correct definitions in the context of the game. Finally, we will investigate the performance of these models in a multi-agent setting where both creativity and logical deduction are crucial for success.

\section{Related Work}
LLMs have demonstrated remarkable success in planning and reasoning capabilities, resulting in the automation of numerous tasks, such as science experiments \citep{zheng2023chatgpt} and software development \citep{qian2023communicative,hong2023metagpt,dong2023self}. The advancement of using an LLM as a planning or decision-making agent has led to significant progress in complex problem-solving and world simulation within LLM-MA systems. One example of world simulation is using memory-based adjustment for LLM agents in games with cooperative or competitive communication paradigms, with either centralized or decentralized communication structures \citep{guo2024large}. For instance, \citet{mukobi2024welfare} use the Welfare game, where LLM agents balance investing in military units and improving their nations' welfare to evaluate the cooperative capabilities of LLMs.

Avalon \citep{wang2023avalons,light2023avalonbench} and Werewolf \citep{xu2024language,xu2023exploring} are two other games used in this paradigm, both with two groups of roles, good and evil, and the winner is the team that succeeds in eliminating the other. The evil group members have the advantage of knowing each other, while the good group members should rely on behavioral patterns to find other members in their group. The most important capabilities examined in these types of games are deceiving other players and distinguishing between the behavioral patterns of good and evil.

\citet{wang2023avalons} compare the performance of Recursive Contemplation (ReCon) and Chain-of-Thought (CoT) \citep{NEURIPS2022_9d560961} for LLM reasoning in the Avalon game, where agents are evaluated by the \texttt{gpt-4-0613} model \citep{openai2024gpt4} using six binary labels (concealment, logic, contribution, persuasiveness, information, and creativity), showing the superiority of ReCon. \citet{light2023avalonbench} also use Avalon for benchmarking LLMs based on their win rate, showing that while LLMs can deduce information from their discussions with other players, they are not able to strategize accordingly. \citet{xu2023exploring} use Werewolf to examine the effect of memory (experience pool) and its size on agent adjustment in the game, where models are shown to improve over rounds based on win rate. \citet{xu2024language} also use the Werewolf game to evaluate LLMs combined with reinforcement learning to examine agent adjustment.

Outside of game simulations, \citet{lee2023natural} and \citet{orwig2024language} use Divergent Thinking (DT) to evaluate LLMs' creativity by calculating semantical differences among multiple responses for a specific topic, e.g., describing a new feasible use case for a typical object. DT is defined as a thought process that enables people to explore and think in multiple directions \citep{guilford1967nature}, which aligns with the objectives of the Balderdash game, explained in the next Section.


\section{The Original Balderdash Game}
\label{sec:balderdash_background}
Balderdash is a word game where players aim to create plausible-sounding definitions for rare and unusual words. The game has two objectives: 1. to deceive other players into believing an invented definition is the correct one, and 2. to correctly identify the true definition among those presented. The game also includes a competitive aspect where players advance on a board towards a finish line.

In each round of the game, the \textit{Dasher} (the leader of each round) draws a card from Balderdash's deck of cards, which contains obscure, rare words along with their definitions. The Dasher announces the chosen word to all players, who then write a definition down on their sheets. Players can either write down the true definition (if they know it) or invent a plausible definition they think will convince others. 

Once all definitions are submitted, the Dasher examines the answers and immediately awards three points to any player whose invented definition closely resembles the true definition. These players do not continue participating in that round. The Dasher then mixes all the remaining invented definitions with the true definition of the word and reads them aloud. Players must vote for the definition they believe is the correct one. Correct guesses are awarded two points, and one point is awarded for each vote a player's definition receives. Additionally, the Dasher receives three points if no player guesses the correct definition. The game continues with a new Dasher each round until one player reaches the finish line on the game board.\footnote{See \url{https://www.hasbro.com/common/instruct/balderdash.pdf} for more detailed instructions.}

\section{LLM-MA Balderdash}
\begin{table*}[t!]
\centering
\begin{tabular}{lccc}
\toprule
\textbf{Model Name} & \textbf{Abbreviation} & \textbf{Reference} & \textbf{\# Parameters} \\
\midrule
meta-llama/Meta-Llama-3-8B-Instruct & Llama & \citet{llama3modelcard} & 8 billion \\
microsoft/Phi-3-small-8k-instruct & Phi & \citet{abdin2024phi3} & 7 billion \\
google/gemma-1.1-7b-it & Gemma & \citet{gemma_2024} & 7 billion \\
mistralai/Mistral-7B-Instruct-v0.3 & Mistral & \citet{jiang2023mistral} & 7 billion \\
gpt-3.5-turbo-0125 & GPT & \citet{openai2024} & Not specified \\
\bottomrule
\end{tabular}
\caption{Summary of the LLMs used in the framework. The models are referenced using their abbreviated names throughout the paper.}
\label{tab:llms}
\end{table*}
We propose a framework where LLMs play the Balderdash game, enabling the benchmarking of their capabilities in generating and evaluating creative content. This framework includes a centralized game engine featuring various LLMs as participants, multiple datasets as the game's word decks, an LLM as the Dasher, and a review of previous rounds given to players as history. These features are discussed in more detail below.

\subsection{LLMs as Participants}
In our framework, LLMs play Balderdash against one another. To incorporate a range of different LLMs, we include four open-source, small, instruct-tuned models loaded locally and one large API-based model (see \autoref{tab:llms}). Each game consists of multiple rounds, with the same set of players participating in each round. Since there are no boards implemented in the framework (and hence no finish line), the players' objective is to maximize their points.

\subsection{Word Deck}
\label{subsec:data}
We created two different datasets used as the Word Decks in our framework. First, we rely on the set of words originally used in the Balderdash game, containing rare and infrequent English words to simulate the actual game. We also created multiple subsets of this dataset containing the words known by each model. According to \citet{kang-choi-2023-impact}, LLMs are biased towards frequent words and co-occurrences, making them vulnerable and unpredictable when infrequent words are used in the input. Therefore, we created another dataset containing the most frequent English words to evaluate LLMs on both frequent and infrequent decks of words. \autoref{tab:dataset} shows a summary of the datasets, along with their average frequency calculated using the NGRAMS Dataset \citep{ngrams}.

\begin{table}[t]
    \centering
    \begin{tabular}{lcc}
        \toprule
        \textbf{Dataset} & \textbf{\# Words} & \textbf{Avg. Frequency} \\ \midrule
        All Balderdash   & 225               & 1.8e-8                  \\
        Llama-Known      & 84                & 3.6e-8                  \\
        Phi-Known        & 88                & 3.7e-8                  \\
        Gemma-Known      & 35                & 5.7e-8                  \\
        Mistral-Known    & 88                & 3.7e-8                  \\
        GPT-Known        & 131               & 2.6e-8                  \\
        Basic English    & 2865              & 6.3e-5                  \\
        \bottomrule
    \end{tabular}
    \caption{Summary of datasets used in this work. The ``Known'' datasets are named using the abbreviated names of the models.}
    \label{tab:dataset}
\end{table}

\subsubsection{Balderdash Words}
We created the ``All Balderdash'' dataset containing 225 distinct Balderdash words sourced from the Wordnik dictionary's list of Balderdash game words\footnote{\url{https://www.wordnik.com/lists/balderdash-game-words}}, complete with all their different definitions and their part of speech tags.

\paragraph{Known Balderdash Words:}
Following \citet{jhirad2023evaluating}, we created datasets of words understood by each LLM by inputting every word along with its part of speech from the ``All Balderdash'' dataset into each model five times, using a temperature value of 0.9. Each model is prompted to act as a universal dictionary and provide a definition of each word. Subsequently, we used \texttt{Llama} as a semantic equivalence judge to determine whether each definition was semantically equivalent to the word's actual definition. We explain this choice in Section~\ref{subsec:llmasjudge}. The prompts provided no context about the Balderdash game (all prompts are detailed in \autoref{sec:appendix_detailed_prompts}). If the model affirmed the semantic equivalence of the majority (three or more out of five) of the definitions, the word is labeled as a ``known'' word for that model.

\subsubsection{Basic Frequent English Words}
We use the Oxford 3000 word list \citep{oxford3000}, containing the most frequent English words. Using the NLTK package \citep{bird-loper-2004-nltk}, English stopwords are removed from this list, resulting in 2895 words. Then, the Merriam-Webster dictionary API \citep{merriamwebsterapi} is used to obtain the various definitions and part of speech tags of these words. Words that do not have any definitions in the Merriam-Webster dictionary are discarded, resulting in 2865 words. The gathered data is cleaned with regular expressions to remove special tokens as defined in the API's documentation. Given that the words in this dataset are the most frequently used English words and thus likely present in the training data of these LLMs, it is expected that even under high-temperature conditions during LLM inference, they will be able to generate accurate definitions for these words.

\subsection{Dasher (Judge)} 
\label{subsec:llmasjudge}
The main responsibility of the Dasher is to act as a judge and examine participants' definitions. Following \citet{zheng2024judging}, where an LLM is used to evaluate open-domain question-answering, we use an LLM as the judge to determine whether each generated definition is semantically equivalent to the reference dictionary definition.

We created a dataset (``Judge Evaluation Data'') to evaluate the best LLM for the Dasher role in the game. This dataset consists of 40 randomly selected words from the ``All Balderdash'' dataset. For each word, \texttt{GPT} was prompted once to provide an accurate definition and again to generate a deceiving definition within the context of the Balderdash game. A human annotator then labeled the \texttt{GPT}-generated definitions (including both correct and deceiving definitions) as ``True'' if they were equivalent to the dictionary definition and ``False'' otherwise. Each LLM was then prompted to do the same task and respond with either ``True'' or ``False.'' The specific prompts used are detailed in \autoref{sec:appendix_detailed_prompts}.

Based on the alignment of human labels and each LLM's labels (see \autoref{tab:judge}), \texttt{Llama} was chosen as the judge of the game in all experiments. Surprisingly, \texttt{GPT} performed the worst. Further investigation revealed that the ``Judge Prompt'' described in Section~\ref{subsec:game_engine} led \texttt{GPT} to become a very strict judge, resulting in generating ``False'' even for small differences in details. We acknowledge that LLMs might have a self-enhancement bias toward their own output or other machine-generated outputs \citep{chen2024humans,zheng2024judging}, resulting in a slightly unfair evaluation. 

It is also worth mentioning that BERTScore \citep{song-etal-2021-sentsim} is another method for calculating semantic distance used in machine translation. However, our experiments detailed in \autoref{sec:appendix_bertscore} demonstrate that it is not feasible to use BERTScore for the judge component.

\begin{table}[t]
    \centering
    \begin{tabular}{lcccc}
        \toprule
        \textbf{LLM} & \textbf{\( F_1 \)} & \textbf{Recall} & \textbf{Precision} & \textbf{Accuracy} \\ \midrule
        Llama        & \textbf{0.74}      & 0.74            & 0.74               & \textbf{0.82}     \\
        Phi          & 0.72               & 0.74            & 0.71               & 0.81              \\
        Gemma        & 0.73               & \textbf{0.77}   & 0.70               & 0.81              \\
        Mistral      & 0.70               & 0.70            & 0.70               & 0.80              \\
        GPT          & 0.19               & 0.11            & \textbf{0.75}      & 0.68              \\
        \bottomrule
    \end{tabular}
    \caption{Evaluation of each LLM as the Dasher using 80 manually labeled data points.}
    \label{tab:judge}
\end{table}

\subsection{History} To provide the players with a memory-based review of previous rounds' outcomes and a sense of their performance, we give each player a history of each round in the form of a CSV file.

\subsection{Game Engine}
\label{subsec:game_engine}
We implemented a game engine capable of simulating Balderdash within a multi-agent environment with centralized communication. In this game engine, LLMs are given five categories of prompts (technical details of the game engine and prompts are available in \autoref{sec:appendix_game_engine_technical_details} and \autoref{sec:appendix_detailed_prompts}):

\paragraph{Game Rules Prompt:} Describes the game rules, scoring rules, and the player's objective, given as a ``system'' prompt. For models that do not support the ``system'' role, this prompt is placed at the beginning of the ``user'' prompt.

\paragraph{History Prompt (Optional):} Provides a review of a moving window of the previous rounds, given as a ``user'' prompt. This approach is to simulate how a human might recall and adapt their strategy over time. The history is available in two versions:
1. Full History includes detailed information for each round, namely round ID, player rank up to that round, score, word, reference definition, generated definition, semantic equivalence, correct guess indicator, deception ratio, and round winners' strategies.
2. Mini History includes a concise version with round ID, player rank up to that round, score, word, and generated definition.

\paragraph{Generate Definition Prompt:} Asks the player to generate a definition for a given word based on the game rules, concatenated with the optional history prompt.

\paragraph{Vote on Definitions Prompt:} Asks the player to choose the reference dictionary definition for a word among all given definitions during the voting phase, concatenated with the optional history prompt.

\paragraph{Judge Prompts:} Consist of a ``system'' and a ``user'' prompt, asking the judge LLM to evaluate whether a reference dictionary definition and a given definition capture the same core concept.

Upon each run, the results of each round are stored in a MongoDB database with collections for games, rounds, and players storing game configurations, player details, and round-specific data:

\paragraph{Games Collection:} Stores overall game configurations, such as game description, number of rounds, judge LLM model name, random seed, scoring rules, history window size, LLMs' temperature, game's word deck, and prompt files used.

\paragraph{Rounds Collection:} Stores round-specific data, including the announced word, its definition, round players' received scores, cast votes, generated definitions, and the judge's evaluation on two aspects: whether each definition is semantically equivalent to the reference definition, and whether it matches at least one of the different meanings of the word (for words with multiple definitions).

\paragraph{Players Collection:} Stores player details, including the LLM name, cumulative score over each round, and rank history in each round.

\section{Evaluation}

Each Balderdash game, denoted as $G_m$, consists of $N$ rounds ($R_n^m$). In each $G_m$, a constant set of $K$ players participate ($P = \{p_1, ..., p_k\}$). The set of all players using the $l^{th}$ LLM is denoted as $LLM_l$. Therefore, each player ($p_k$) is a member of one and only one $LLM_l$. $R_n^m$ contains information about all players participating in the $n^{th}$ round of $G_m$, including ``\texttt{judge decision}'', ``\texttt{llm knows one}'', ``\texttt{votes}'', and ``\texttt{scores}''. The first two are mappings between each $p_k$ and a binary value, indicating whether $p_k$'s generated definition was semantically equal to the first reference dictionary definition of $R_n^m$'s word and whether $p_k$'s output was semantically equal to at least one of the various definitions of $R_n^m$'s word, respectively. ``\texttt{votes}'' is another mapping containing information on each $p_k$'s vote in the voting phase, either for another player ($p_{k'}$) or for ``$-1$'', representing the reference dictionary definition. ``\texttt{scores}'' is a mapping between each $p_k$ and an integer value indicating $p_k$'s score in $R_n^m$.

\subsection{Metrics}
We define five metrics for each round ($R_n^m$): 1. True Definition Ratio (TDR), 2. LLM Knows Ratio (LKR), 3. Deception Ratio (DR), 4. Correct Guess Ratio (CGR), and 5. Average Score (AS). $TDR_n^m(LLM_l)$ represents the ratio of true definitions generated for the announced word in the $m^{th}$ game and the $n^{th}$ round for all players in $LLM_l$. 
\begin{equation}
    \begin{split}
        T&DR_n^m(LLM_l) = \\
        &\frac{\sum_{p_k \in LLM_l} R_{n}^{m}(\texttt{judge decision})[p_k]}{|LLM_l|}
    \end{split}
\end{equation}
LKR measures the ratio of instances where the LLM aims to generate the true definition. 
\begin{equation}
    \begin{split}
        L&KR_n^m(LLM_l) = \\
        &\frac{\sum_{p_k \in LLM_l} R_{n}^{m}(\texttt{llm knows one})[p_k]}{|LLM_l|}
    \end{split}
\end{equation}
The metrics DR and CGR are designed to evaluate the performance of each LLM in the voting phase. DR measures the success ratio of LLMs in deceiving other players.
\begin{equation}
    \begin{split}
        D&R_n^m(LLM_l) = \\
        &\frac{1}{|LLM_l|}\sum_{p_k \in LLM_l} \frac{\sum_{v \in R_n^m(\texttt{votes})}\delta(v,p_k)}{|R_n^m (\texttt{votes})|-1}
    \end{split}
\end{equation}
CGR reflects the LLMs' ability to identify the reference dictionary definition amidst deceiving ones. 
\begin{equation}
    \begin{split}
        C&GR_n^m(LLM_l) = \\
        &\frac{\sum_{p_k \in LLM_l} \delta(R_n^m(\texttt{votes})[p_k], -1)}{|LLM_l|}
    \end{split}
\end{equation}
AS is the average score achieved by an LLM. This metric also represents a weighted summation of TDR, DR, and CGR, where the weights are determined by the game's scoring rules.
\begin{equation}
    \begin{split}
        A&S_n^m(LLM_l) = \\
        &\frac{\sum_{p_k \in LLM_l} R_{n}^{m}(\texttt{scores})[p_k]}{|LLM_l|}
    \end{split}
\end{equation}
   
The above metrics are used to assess the overall performance of LLMs in the LLM-MA Balderdash game. In cases where there is a dominant strategy that allows players to get the most points, such as generating the correct definition when the correct definition score is set to a high value, we define convergence to assess the LLM's strategy. The goal of convergence is to determine if the model can find and continuously use the most rewarding strategy. Convergence is defined as follows:

\begin{equation}
    \overline{LKR}_n > 1 - \epsilon, \quad \forall n > T
\end{equation}

\section{Experiments \& Results}

To evaluate the LLMs' performance and strategy, we conduct three experiments. The first experiment provides a leaderboard of LLMs based on their proficiency in playing the original Balderdash game. The second experiment investigates whether LLMs learn from their history and converge to follow the most rewarding strategy. The final experiment targets LLMs' ability to reason over game rules and choose the best greedy choices.

\begin{table*}[t!]
\centering
\begin{tabular}{|l|l|c|c|c|c|c|}
\toprule
\textbf{HT}         & \textbf{LLM}                                 & \textbf{LKR} & \textbf{TDR} & \textbf{DR} & \textbf{CGR} & \textbf{AS} \\ \midrule
\multirow{4}{*}{\rotatebox[origin=c]{90}{none}} & Llama & 0.59 ± 0.10       & 0.40 ± 0.04       & 0.19 ± 0.12       & 0.56 ± 0.11       & 2.08 ± 0.19    \\  
                       & Phi   & 0.49 ± 0.12       & 0.33 ± 0.10       & 0.24 ± 0.13       & \textbf{0.77 ± 0.12}       & 2.21 ± 0.30    \\  
                       & Gemma & \textbf{0.93 ± 0.02}       & \textbf{0.77 ± 0.08}       & \textbf{0.25 ± 0.13}       & 0.47 ± 0.19       & \textbf{2.65 ± 0.13}    \\  
                       & Mistral & 0.59 ± 0.13       & 0.48 ± 0.17       & 0.15 ± 0.07       & 0.74 ± 0.13       & 2.35 ± 0.29    \\ \midrule
\multirow{4}{*}{\rotatebox[origin=c]{90}{mini}} & Llama & 0.89 ± 0.12       & 0.78 ± 0.14       & 0.28 ± 0.18       & 0.49 ± 0.20       & 2.68 ± 0.15    \\  
                       & Phi   & 0.74 ± 0.18       & 0.60 ± 0.18       & 0.30 ± 0.17       & \textbf{0.74 ± 0.07}       & 2.52 ± 0.22    \\  
                       & Gemma & 0.92 ± 0.09       & 0.78 ± 0.12       & 0.30 ± 0.28       & 0.43 ± 0.16       & 2.63 ± 0.21    \\ 
                       & Mistral & \textbf{0.94 ± 0.07}       & \textbf{0.83 ± 0.07}       & \textbf{0.34 ± 0.18}       & 0.35 ± 0.32       & \textbf{2.76 ± 0.06}    \\ \midrule
\multirow{4}{*}{\rotatebox[origin=c]{90}{full}} & Llama & 0.93 ± 0.06       & 0.78 ± 0.16       & 0.61 ± 0.07       & \textbf{0.52 ± 0.31}       & 2.72 ± 0.18    \\  
                       & Phi   & 0.98 ± 0.04       & 0.85 ± 0.05       & 0.53 ± 0.34       & 0.42 ± 0.26       & 2.79 ± 0.07    \\  
                       & Gemma & 0.97 ± 0.04       & 0.84 ± 0.07       & 0.52 ± 0.29       & 0.44 ± 0.34       & 2.69 ± 0.21    \\  
                       & Mistral & \textbf{1.00 ± 0.00}       & \textbf{0.90 ± 0.05}       & \textbf{0.62 ± 0.35}       & 0.05 ± 0.10       & \textbf{2.81 ± 0.06}    \\ \bottomrule
\end{tabular}
\caption{Leaderboard experiment results on ``Basic Frequent English Words,'' evaluating each LLM in three different settings based on history type (HT) using the average of LKR, TDR, DR, CGR, and AS metrics over all rounds and games. The highest value of each metric for different game settings is in bold. However, based on the standard deviation, this does not represent absolute superiority.}
\label{tab:leaderboard_basic}
\end{table*}

\subsection{Leaderboard Experiment}
In this experiment, we aim to create a leaderboard for LLMs by having these models play Balderdash against each other. To keep the game fair, only models of comparable size (namely \texttt{Llama}, \texttt{Phi}, \texttt{Gemma}, and \texttt{Mistral}) are used. Using more advanced models would disrupt the game flow, as smaller models wouldn't be able to rise in the rankings and consequently learn from their history. Each game with four players representing four LLMs is run five times using five different subsets of words to ensure that the chosen set of words does not affect the results. This experiment is conducted with three types of history (none, mini, and full) and two datasets (``Basic Frequent English Words'' and ``All Balderdash'') to examine the models' performance on both frequent and infrequent English words.

The results for ``Basic Frequent English Words,'' shown in \autoref{tab:leaderboard_basic}, indicate a considerable improvement for all models as the history becomes more informative. The only metric that decreases is CGR. As LKR approaches $1.0$ for all models with increasing history, the ratio of rounds with more than one correct definition in the voting phase also increases. This could lead to confusion for all players and possibly result in a drop in CGR because the definitions in the voting phase are true definitions of the word but not the reference one used by the judge.

\begin{table*}[t]
\centering
\begin{tabular}{|l|l|c|c|c|c|c|}
\toprule
\textbf{HT}         & \textbf{LLM}                                 & \textbf{LKR} & \textbf{TDR} & \textbf{DR} & \textbf{CGR} & \textbf{AS} \\ \midrule
\multirow{4}{*}{\rotatebox[origin=c]{90}{none}} & Llama & 0.27 ± 0.09       & 0.22 ± 0.11       & 0.26 ± 0.06       & 0.57 ± 0.12       & 2.00 ± 0.25    \\  
                       & Phi   & 0.31 ± 0.13       & 0.25 ± 0.11       & 0.19 ± 0.05       & \textbf{0.62 ± 0.05}       & \textbf{2.08 ± 0.22}    \\  
                       & Gemma & 0.18 ± 0.10       & 0.15 ± 0.08       & 0.12 ± 0.03       & 0.35 ± 0.08       & 1.26 ± 0.15    \\  
                       & Mistral & \textbf{0.44 ± 0.17}       & \textbf{0.33 ± 0.17}       & \textbf{0.27 ± 0.04}       & 0.45 ± 0.08       & 2.05 ± 0.28    \\ \midrule
\multirow{4}{*}{\rotatebox[origin=c]{90}{mini}} & Llama & 0.29 ± 0.15       & 0.24 ± 0.12       & \textbf{0.30 ± 0.07}       & 0.54 ± 0.12       & 2.04 ± 0.17    \\  
                       & Phi   & \textbf{0.45 ± 0.16}       & \textbf{0.35 ± 0.15}       & 0.26 ± 0.09       & \textbf{0.56 ± 0.13}       & \textbf{2.28 ± 0.19}    \\  
                       & Gemma & 0.07 ± 0.07       & 0.05 ± 0.05       & 0.10 ± 0.06       & 0.32 ± 0.12       & 0.96 ± 0.35    \\  
                       & Mistral & 0.44 ± 0.17       & \textbf{0.35 ± 0.13}       & \textbf{0.30 ± 0.07}       & 0.41 ± 0.11       & 2.05 ± 0.25    \\ \midrule 
\multirow{4}{*}{\rotatebox[origin=c]{90}{full}} & Llama & 0.33 ± 0.16       & 0.24 ± 0.15       & 0.20 ± 0.06       & 0.39 ± 0.12       & 1.70 ± 0.27    \\  
                       & Phi   & \textbf{0.40 ± 0.14}       & \textbf{0.37 ± 0.14}       & \textbf{0.33 ± 0.08}       & \textbf{0.66 ± 0.13}       & \textbf{2.52 ± 0.25}    \\  
                       & Gemma & 0.17 ± 0.10       & 0.13 ± 0.10       & 0.19 ± 0.09       & 0.26 ± 0.13       & 1.19 ± 0.48    \\  
                       & Mistral & 0.36 ± 0.18       & 0.31 ± 0.15       & 0.28 ± 0.06       & 0.36 ± 0.12       & 1.89 ± 0.27    \\ \bottomrule
\end{tabular}
\caption{Leaderboard experiment results on ``All Balderdash,'' evaluating each LLM in three different settings based on history type (HT) using the average of LKR, TDR, DR, CGR, and AS metrics over all rounds and games. The highest value of each metric for different game settings is in bold. However, based on the standard deviation, this does not represent absolute superiority.}
\label{tab:leaderboard_balderdash}
\end{table*}

The results for ``All Balderdash'' are shown in \autoref{tab:leaderboard_balderdash}. Contrary to the ``Basic Frequent English Words'' results, consistent improvement is not observed for all LLMs. A possible reason could be the infrequency of the words in this dataset. In almost all settings, \texttt{Phi} performs strongly in finding the correct definition during the voting phase, suggesting its potential for detecting disinformation. Furthermore, \texttt{Mistral} shows the best overall performance in deceiving its opponents, possibly due to greater creativity in generating deceptive definitions, as deception in Balderdash is an iterative process requiring creativity to avoid pattern recognition.

None of the models dominate the others across all game settings. However, when using the ``Basic Frequent English Words'' dataset, \texttt{Mistral} has the most wins, whereas when using the ``All Balderdash'' dataset, \texttt{Phi} performs best overall. It is worth mentioning that \texttt{Mistral} was the only model that failed to conform to the specified format in the voting prompt in two games.

\subsection{Convergence Experiment}
Although the leaderboard provides some insight into LLMs' performance, evaluating their strategies and understanding their behavior remains challenging. Therefore, this experiment aims to evaluate LLMs' reasoning and strategy in an environment where a dominant method for maximizing scores exists based on the history of past rounds. The dataset used in this experiment is limited to ``Known Balderdash Words'' for each LLM, and the game is run with three players using the same LLM (including \texttt{GPT}). Considering that the players know the definitions of the announced words, we hypothesize that in each game, the LLMs' LKR will converge since generating the true definition is the most rewarding strategy. Similar to the first experiment, each game is run five times with five different subsets of the dataset. Only two types of history (mini and full) are used in this experiment.

\begin{figure*}[]
  \centering
    \includegraphics[width=\linewidth]{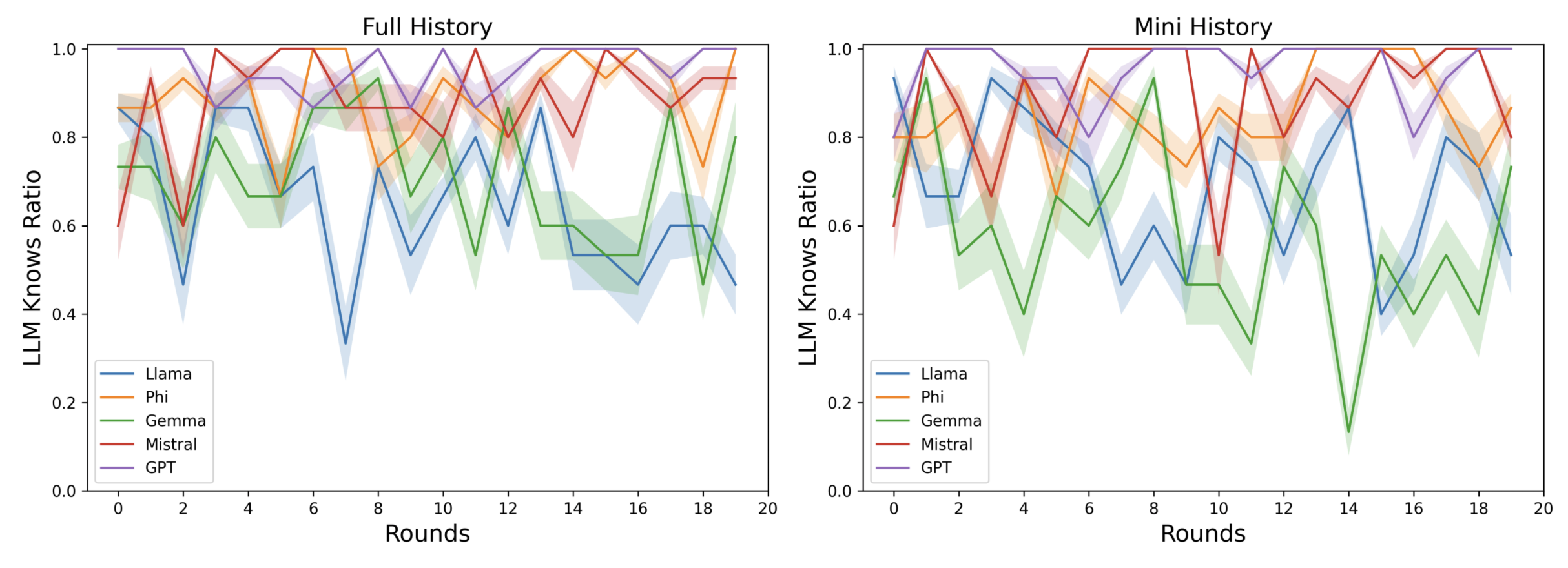} 
  \caption{Convergence experiment on ``Known Balderdash Words'' with mini and full history types, examining changes in $\overline{LKR}_n$ over rounds. Note that for all LLMs, the standard deviation is down-scaled by a factor of 0.2 for presentation purposes.}
  \label{figexp2}
\end{figure*}

\autoref{figexp2} depicts $\overline{LKR}_n$ over rounds, showing that none of the models converge, contrary to our hypothesis. The plots show a reduction in fluctuations for the full history setting compared to the mini history, but still, there is no improvement or trend for any of the models over rounds. This phenomenon could be due to the infrequency of the words in the dataset or a weakness of these LLMs in finding or repeatedly using the best strategy.

\subsection{Game Rules Experiment}
The final experiment aims to assess LLMs' ability to understand and reason over the game rules without providing history. This experiment is conducted with one player, using the ``Known Balderdash Words'' dataset for each LLM, and two distinct rule sets: 1. awarding fifty points for generating the true definition, and 2. awarding zero points for the same task, both with the same scoring rules for correct guesses and receiving votes from other players in the voting phase. In this setting, our hypothesis is that LLMs will choose the most rewarding strategy in each setting in a greedy manner, which is generating the true definition and guessing the correct definition in the first and second experiment settings, respectively.

To assess our hypothesis, $\overline{TDR}$ and $\overline{LKR}$ were calculated in both settings (\autoref{tabexp3}). Although there is a slight increase in $\overline{TDR}$ for \texttt{Mistral} and \texttt{GPT}, the results are still disappointing. Even with zero points for generating the true definition, models are still choosing this strategy, leading to zero points in each round.

\begin{table*}[t]
\centering
\begin{tabular}{lcccc}
\toprule
\textbf{LLM}                 & \multicolumn{2}{c}{\textbf{LKR}} & \multicolumn{2}{c}{\textbf{TDR}} \\ \hline
\textbf{Correct Def. Points} & \multicolumn{1}{c}{\textbf{0}}  & \textbf{50}  & \multicolumn{1}{c}{\textbf{0}}          & \textbf{50}  \\ \hline
Llama     & \multicolumn{1}{c}{0.60 ± 0.04}  & 0.59 ± 0.08  & \multicolumn{1}{c}{0.55 ± 0.04} & 0.49 ± 0.09 \\ 
Phi      & \multicolumn{1}{c}{0.64 ± 0.08}  & 0.62 ± 0.12 & \multicolumn{1}{c}{0.55 ± 0.07} & 0.54 ± 0.09 \\ 
Gemma              & \multicolumn{1}{c}{0.53 ± 0.09}   & 0.50 ± 0.07 & \multicolumn{1}{c}{0.46 ± 0.14} & 0.46 ± 0.08 \\ 
Mistral     & \multicolumn{1}{c}{0.79 ± 0.09}  & 0.82 ± 0.09 & \multicolumn{1}{c}{0.65 ± 0.08} & 0.68 ± 0.11 \\ 
GPT           & \multicolumn{1}{c}{0.92 ± 0.04}  & 0.93 ± 0.07 & \multicolumn{1}{c}{0.86 ± 0.04} & 0.88 ± 0.05 \\ \bottomrule
\end{tabular}
\caption{Examining LLM reasoning through the effect of game rules.}
\label{tabexp3}
\end{table*}

\section{Conclusion}
Current LLM-MA game simulations overlook the assessment of creativity in LLMs. This study introduces a systematic framework through the Balderdash game to probe aspects of creativity, deception, and logical reasoning inherent in these models. Our initial assumption was that LLMs are familiar with most Balderdash words and can learn the patterns in machine-generated deceiving definitions, thereby enabling them to generate the correct definitions of words and choose the dictionary definition in the voting phase of Balderdash.

Contrary to our expectations, LLMs are not familiar with more than half of the Balderdash words and perform poorly during the voting phase. None of the models used in the experiments showed signs of correct reasoning based on game rules or strategy convergence derived from historical context. Interestingly, this phenomenon is more pronounced with Balderdash words (infrequent English words) compared to more frequent English words, suggesting that LLMs are more susceptible to failure in reasoning when encountering infrequent vocabulary.

The best judge among all LLMs we tested was \texttt{Llama}, which had the best alignment with human labels. Based on the leaderboard experiment, \texttt{Phi} performed strongly in finding the correct definition during the voting phase, suggesting its potential for detecting disinformation. Furthermore, \texttt{Mistral} showed the best overall performance in deceiving its opponents, likely due to its creativity in generating deceptive definitions.

\section*{Limitations}
The judge in the LLM-MA Balderdash plays a crucial role in both running the game and evaluating its outcomes. Consequently, the accuracy of the judge is a critical factor in our work. In the current version of the game engine, an LLM serves as the judge. However, an alternative could involve replacing the LLM judge with a specialized model specifically trained to discriminate between true and deceiving definitions. This replacement would likely result in higher accuracy and a more reliable game simulation system.

Additionally, the possibility of self-enhancement bias should be considered when using an LLM as the judge. To evaluate this bias, we can assess each LLM as the judge on definitions generated not only by \texttt{GPT} (the model we’re using) but also by all other LLMs employed in our work. By comparing error rates across different sets of generated definitions, we can gain insights into how biased these models are toward their own output.

We create subsets of words from the ``All Balderdash'' dataset understood by each LLM by probing the models' output using a prompt to generate a definition for each word. Given the low frequency of these words and the high temperature value in our setting, this might lead to false negatives. A better method would involve comparing the frequency of occurrence of ``All Balderdash'' words to that of ``Basic Frequent English Words'' in each LLM's training data (if available).

Currently, we use a high temperature during inference to create diversity in generated responses. Alternatively, using temperature scaling or diverse prompting methods for each player might result in more reliable outputs, especially for infrequent words. Furthermore, all evaluations of LLMs' strategy and reasoning in this work are based on game results and scores. Replacing current prompts with multi-stage reasoning and Chain-of-Thought (CoT) approaches might improve LLMs' performance in the game and provide better insights into the reasoning behind their strategies.



\bibliography{main}

\appendix

\section{BERTScore as a Judge}
\label{sec:appendix_bertscore}

Following \citet{song-etal-2021-sentsim}, where semantic distance is used for evaluating machine translation, BERTScore is calculated between all true definitions of each word in the ``Judge Evaluation Data'' and its respective deceiving definition. The maximum $F_1$-score of the calculated BERTScores for each word is used. For 60\% of the words in the ``Judge Evaluation Data,'' the maximum $F_1$-score of the true definition is higher than that of the deceiving definition. However, it is not possible to define a clear threshold for semantic equivalence using BERTScore, as scores for both true and deceiving definitions overlap, ranging between 0.8 and 0.9. Therefore, it was not possible to use BERTScore output for our judge component.

\section{Computational Resources}
All the experiments were conducted on three \texttt{NVIDIA RTX A6000} GPUs with 48GB VRAM. In total, all the experiments took around 8 hours to complete.

\section{Game Engine Technical Details}
\label{sec:appendix_game_engine_technical_details}
This platform supports using different sets of LLMs, each with different chat templates, GPU mappings, and parameters such as temperature or maximum number of new tokens to generate. These LLMs can be assigned as individual players or the game judge, or a single LLM can represent all game roles.

We use Huggingface's transformers \citep{wolf-etal-2020-transformers} and accelerate \citep{accelerate} libraries to load all local, open-source models in our game engine. This allows for loading variations of these models, such as different quantizations, or completely changing them by simply modifying the model names in the game engine configuration. For API-based models, such as \texttt{GPT}, we use OpenAI's python library. Other OpenAI models can be used by changing the model names in the game engine configurations.

\section{Detailed Prompts}
\label{sec:appendix_detailed_prompts}

\begin{figure}[h]
\centering
\begin{tcolorbox}[colback=gray!10, colframe=black, title=User Prompt, boxrule=1pt, arc=2mm, outer arc=2mm, left=1mm, right=1mm]
\texttt{define the word \{word\} in one sentence.}
\end{tcolorbox}
\caption{The prompt for generating a correct definition of words to create the ``Judge Evaluation Data''.}
\end{figure}

\begin{figure}[h]
\centering
\begin{tcolorbox}[colback=gray!10, colframe=black, title=User Prompt, boxrule=1pt, arc=2mm, outer arc=2mm, left=1mm, right=1mm]
\texttt{Suppose you are playing Balderdash, and you want to deceive your opponents with a deceiving definition. define \{word\} in one sentence.}
\end{tcolorbox}
\caption{The prompt for generating deceiving definitions of words to create the ``Judge Evaluation Data''.}
\end{figure}

\begin{figure}[h]
\centering
\begin{tcolorbox}[colback=gray!10, colframe=black, title=System Prompt, boxrule=1pt, arc=2mm, outer arc=2mm, left=1mm, right=1mm]
\texttt{You are a universal dictionary. Your task is to provide the definition of a word given to you along with its part of speech. Use at most one sentence.}
\end{tcolorbox}
\begin{tcolorbox}[colback=gray!10, colframe=black, title=User Prompt, boxrule=1pt, arc=2mm, outer arc=2mm, left=1mm, right=1mm]
\texttt{\{word\} (\{pos\}): }
\end{tcolorbox}
\caption{The prompts for generating definitions to create the ``Known Balderdash Words'' dataset.}
\end{figure}

\begin{figure}[h]
\centering
\begin{tcolorbox}[colback=gray!10, colframe=black, title=System Prompt, boxrule=1pt, arc=2mm, outer arc=2mm, left=1mm, right=1mm]
\texttt{You are a semantical equivalence judge. Your task is to determine whether a given definition is semantically equal to the actual definition of a word. Give your answer as a single word, either 'true' or 'false'.}
\end{tcolorbox}
\begin{tcolorbox}[colback=gray!10, colframe=black, title=User Prompt, boxrule=1pt, arc=2mm, outer arc=2mm, left=1mm, right=1mm]
\texttt{Actual definition: \{definition\} and generated definition: \{llm\_definition\}. Your judgment:}
\end{tcolorbox}
\caption{The prompts for semantical equivalence judgment used to create the ``Known Balderdash Words'' dataset.}
\end{figure}

\begin{figure*}[ht]
\centering
\begin{tcolorbox}[colback=gray!10, colframe=black, title=Judge - System Prompt, boxrule=1pt, arc=2mm, outer arc=2mm, left=1mm, right=1mm]
\texttt{Please act as a judge and evaluate whether a reference dictionary definition of a word and a given definition by an assistant capture the same core concept. Consider the essence and key components in the reference dictionary definition. You will be given a word, its reference dictionary definition, and an assistant's definition of that word. Your job is to assess if the assistant's definition captures the essential meaning or core concept of the reference dictionary definition, even if there are minor differences in detail or scope. Begin your evaluation by comparing the assistant's definition with the reference dictionary definition. Maintain objectivity and ensure that the order in which the definitions were presented does not influence your decision. Do not allow the length of the definitions to influence your evaluation. Output your final verdict by strictly following this format: "true" if the assistant's definition captures the core concept of the reference dictionary definition, and "false" if the assistant's definition fails to capture the core concept of the reference dictionary definition.}
\end{tcolorbox}
\begin{tcolorbox}[colback=gray!10, colframe=black, title=Judge - User Prompt, boxrule=1pt, arc=2mm, outer arc=2mm, left=1mm, right=1mm]
\texttt{You receive the word: "\{word\}", the reference dictionary definition: "\{correct\_definition\}", and assistant's definition: "\{definition\}". Give your answer as a single word, either "true" or "false". Your verdict is: }
\end{tcolorbox}
\caption{The prompts for the semantical equivalence judge, used in the game engine.}
\end{figure*}

\begin{figure*}[ht]
\centering
\begin{tcolorbox}[colback=gray!10, colframe=black, title=Generate Definition - User Prompt, boxrule=1pt, arc=2mm, outer arc=2mm, left=1mm, right=1mm]
\texttt{Please act as a competitive player in the Balderdash game and write a definition for the word "\{word\}". You should consider the game's scoring rules and the history of your performance in previous rounds. Begin by elaborating on this information and choose the strategy for writing your definition to maximize your points. Avoid explaining your strategy and only write your definition. Avoid rewriting the given word at the beginning of your definition. Use at most one sentence. Output your definition between double quotes strictly following this format.}
\end{tcolorbox}
\begin{tcolorbox}[colback=gray!10, colframe=black, title=Generate Definition (No History Version) - User Prompt, boxrule=1pt, arc=2mm, outer arc=2mm, left=1mm, right=1mm]
\texttt{Please act as a competitive player in the Balderdash game and write a definition for the word "\{word\}". You should consider the game's scoring rules. Begin by elaborating on this information and choose the strategy for writing your definition to maximize your points. Avoid explaining your strategy and only write your definition. Avoid rewriting the given word at the beginning of your definition. Use at most one sentence. Output your definition between double quotes strictly following this format.}
\end{tcolorbox}
\caption{The generate definition prompt, in two versions for asking the player to generate a definition for a given word based on the game rules, concatenated with the optional history prompt.}
\end{figure*}

\begin{figure*}[ht]
\centering
\begin{tcolorbox}[colback=gray!10, colframe=black, title=Full History - User Prompt, boxrule=1pt, arc=2mm, outer arc=2mm, left=1mm, right=1mm]
\texttt{Your performance history in the previous rounds will be provided, helping you to learn from past performances, better understand your opponents' behaviors, and adapt your strategies to maximize your scoring potential in future rounds. History is provided in CSV format between triple backticks. Columns descriptions of the CSV:
round\_id: The id for the corresponding round.
rank\_among\_players: An integer indicating your rank among all players up to that round.
score: An integer indicating your score in that round.
word: The announced word in that round.
definition: The reference dictionary definition of the announced word.
generated\_definition: Your definition for the announced word.
wrote\_true\_definition: A boolean showing whether the reference dictionary definition of your definition captures the same core concept. If the value of this column is True, you have not participated in the voting phase on that round, and thus, the `guessed\_correct\_definiton` column will be False.
guessed\_correct\_definiton: A boolean showing whether you have correctly guessed the reference dictionary definition in the voting phase.
deception\_ratio: The ratio of players who voted to your definition excluding yourself in the voting phase divided by the total number of players who participated in the voting phase. If the `wrote\_true\_definition` is True, then this value will be -1.
round\_winners\_strategies: A list of tuples containing the definition and that definition's outcome for each of the player(s) who got the highest scores in the corresponding round, in the format of [(definition\_round\_id, outcome\_for\_definition\_round\_id)].
```
\{history\_csv\}
```}
\end{tcolorbox}
\begin{tcolorbox}[colback=gray!10, colframe=black, title=Mini History - User Prompt, boxrule=1pt, arc=2mm, outer arc=2mm, left=1mm, right=1mm]
\texttt{Your performance history in the previous rounds will be provided, helping you to learn from past performances, better understand your opponents' behaviors, and adapt your strategies to maximize your scoring potential in future rounds. History is provided in CSV format between triple backticks. Columns descriptions of the CSV:
round\_id: The id for the corresponding round.
rank\_among\_players: An integer indicating your rank among all players up to that round.
score: An integer indicating your score in that round.
word: The announced word in that round.
generated\_definition: Your definition for the announced word.
```
\{history\_csv\}
```}
\end{tcolorbox}
\caption{The full and mini history prompts, used for providing the performance history of each player in the game engine.}
\end{figure*}

\begin{figure*}[ht]
\centering
\begin{tcolorbox}[colback=gray!10, colframe=black, title=Game Rules - System Prompt, boxrule=1pt, arc=2mm, outer arc=2mm, left=1mm, right=1mm]
\texttt{Please act as a competitive player in the Balderdash game. In each round of the game, a rare and unusual word will be given to all players. The players then write down a definition, which may be an honest attempt to supply the reference dictionary definition or, if they do not know or, for tactical reasons, decide not to, a fictitious definition for the word designed to sound convincing. Players submitting a definition that is semantically equal to the reference dictionary definition are immediately awarded \{correct\_definition\_points\} points, and they will not continue playing on that round. Then, the remaining definitions, including the reference dictionary definition, are given to each player in random order. Then, players write which definition they believe is the reference dictionary definition. Players are awarded \{correct\_vote\_points\} points if they guess the correct definition. Players are awarded \{receiving\_vote\_points\} points for each other player who incorrectly chooses the fake definition they wrote. Your goal is to maximize your points in each round by selecting the best strategy in writing a definition for the word and in the voting phase. You will be given a history of the previous rounds. Use the information in the history and pay attention to the scoring rules to choose the best strategy.}
\end{tcolorbox}
\begin{tcolorbox}[colback=gray!10, colframe=black, title=Game Rules (No History Version) - System Prompt, boxrule=1pt, arc=2mm, outer arc=2mm, left=1mm, right=1mm]
\texttt{Please act as a competitive player in the Balderdash game. In each round of the game, a rare and unusual word will be given to all players. The players then write down a definition, which may be an honest attempt to supply the reference dictionary definition or, if they do not know or, for tactical reasons, decide not to, a fictitious definition for the word designed to sound convincing. Players submitting a definition that is semantically equal to the reference dictionary definition are immediately awarded \{correct\_definition\_points\} points, and they will not continue playing on that round. Then, the remaining definitions, including the reference dictionary definition, are given to each player in random order. Then, players write which definition they believe is the reference dictionary definition. Players are awarded \{correct\_vote\_points\} points if they guess the correct definition. Players are awarded \{receiving\_vote\_points\} points for each other player who incorrectly chooses the fake definition they wrote. Your goal is to maximize your points in each round by selecting the best strategy in writing a definition for the word and in the voting phase. Pay attention to the scoring rules to choose the best strategy.}
\end{tcolorbox}
\caption{The game rules prompt, in two versions for providing instructions on playing the game and the game's scoring rules to each player used in the game engine.}
\end{figure*}

\begin{figure*}[ht]
\centering
\begin{tcolorbox}[colback=gray!10, colframe=black, title=Vote on Definitions - User Prompt, boxrule=1pt, arc=2mm, outer arc=2mm, left=1mm, right=1mm]
\texttt{Please act as a competitive player in the Balderdash game and choose the reference dictionary definition index. You will be given the word for this round, your given definition, and the other definitions, excluding the definitions that were semantically equal to the reference dictionary definition, including the reference dictionary definition in random order. You should consider the game's scoring rules and the history of your performance in previous rounds. Begin by elaborating on this information and choose the reference dictionary definition, which will maximize your points. Avoid explaining your strategy. Choose your vote among the allowed choice(s) and only write your vote. Your definition for "\{word\}" was "\{definition\}". All definitions, including the reference dictionary definition, are given to you in the format:
1. definition\_1
2. definition\_2
3. definition\_3
Definitions:
```
\{definitions\}
```
As a Balderdash player, your task is to choose the reference dictionary definition index \{all\_indexes\_excluding\_player\_descriptive\} and write it without any explanation.
Your allowed choice(s): \{all\_indexes\_excluding\_player\}
Use at most one character, which is a single digit.}
\end{tcolorbox}
\begin{tcolorbox}[colback=gray!10, colframe=black, title=Vote on Definitions (No History Version) - User Prompt, boxrule=1pt, arc=2mm, outer arc=2mm, left=1mm, right=1mm]
\texttt{Please act as a competitive player in the Balderdash game and choose the reference dictionary definition index. You will be given the word for this round, your given definition, and the other definitions, excluding the definitions that were semantically equal to the reference dictionary definition, including the reference dictionary definition in random order. You should consider the game's scoring rules. Begin by elaborating on this information and choose the reference dictionary definition, which will maximize your points. Avoid explaining your strategy. Choose your vote among the allowed choice(s) and only write your vote. Your definition for "\{word\}" was "\{definition\}". All definitions, including the reference dictionary definition, are given to you in the format:
1. definition\_1
2. definition\_2
3. definition\_3
Definitions:
```
\{definitions\}
```
As a Balderdash player, your task is to choose the reference dictionary definition index \{all\_indexes\_excluding\_player\_descriptive\} and write it without any explanation.
Your allowed choice(s): \{all\_indexes\_excluding\_player\}
Use at most one character, which is a single digit.}
\end{tcolorbox}
\caption{The vote on definitions prompt, in two versions for asking the player to choose the reference dictionary definition for a word among all given definitions during the voting phase, concatenated with the optional history prompt.}
\end{figure*}


\end{document}